\documentclass[aps,pra,showpacs,groupedaddress,superscriptaddress,twocolumn]{revtex4-1}

\usepackage{amssymb}
\usepackage{graphicx}
\usepackage{amsmath}
\usepackage{subfigure}
\usepackage{float}
\usepackage{multirow}
\usepackage{mathrsfs}

\newcommand{\bear}{\begin{eqnarray}}
\newcommand{\eear}{\end{eqnarray}}
\newcommand{\be}{\begin{equation}}
\newcommand{\ee}{\end{equation}}
\newcommand{\beqn}{\begin{eqnarray}}
\newcommand{\eeqn}{\end{eqnarray}}
\newcommand{\beqnn}{\begin{eqnarray*}}
\newcommand{\eeqnn}{\end{eqnarray*}}

\begin{document}

\title{On the paradoxical evolution of the number of photons in a new model
of interpolating Hamiltonians}
\author{C.~Valverde}
\email{valverde@ueg.br}
\affiliation{Campus de Ci\^{e}ncias Exatas e Tecnol\'ogicas, Universidade
Estadual de Goi\'{a}s, 75001-970 An\'{a}polis, GO, Brazil} 
\affiliation{Universidade Paulista, 74845-090 Goi\^{a}nia, GO,
Brazil}

\author{B.~Baseia}
\affiliation{Instituto de F\'{\i}sica, Universidade Federal de Goi\'as,
74.690-900 Goi\^ania, GO, Brazil} 
\affiliation{Departamento de
F\'{\i}sica, Universidade Federal da Para\'{\i}ba,  58.051-970 Jo\~{a}o
Pessoa, PB, Brazil}
\date{\today }

\begin{abstract}
We introduce a new Hamiltonian model which interpolates between the
Jaynes-Cummings model and other types of such Hamiltonians. It works with
two interpolating parameters, rather than one as traditional. Taking
advantage of this greater degree of freedom, we can perform continuous
interpolation between the various types of these Hamiltonians. As
applications we discuss a paradox raised in literature and compare the time
evolution of photon statistics obtained in the various interpolating models.
The role played by the average excitation in these comparisons is also
highlighted.
\end{abstract}

\maketitle

\section{Introduction}

The Jaynes-Cummings model (JCM), proposed in 1963 \cite{JCM}, constitutes an
excellent theoretical approach to describe analytically the interaction of a
two level atom with a single mode of a quantized radiation field. The field
frequency may belong either to the optical domain or to the microwave one.
In the first case the researchers use common atoms \cite{a1} whereas in
second case they use (highly excited) Rydberg atoms \cite{ab}. The issue was
also extended to other systems, as (i) in nanocircuits operating in
microwave domain, either through the substitution of the atom by a
Copper-pair box (CPB) and the field by a nanomechanical resonator in
nanocavities \cite{ac}; (ii) or the CPB inside a chip \cite{ac1,ac2};\ (iii)
substituting the atom by quantum a dot embedded in a photonic-crystal \cite%
{ad}; (iv) using spin in quantum-dot arrays \cite{ae}, etc. In spite of its
simplicity the JCM gives exact solutions of the Schr\"{o}dinger equation in
many examples that occur in such physical systems.

The JCM has been employed in the study of various fundamental quantum
aspects involving the matter-radiation. To give some examples we mention:
collapse and revival of the atomic inversion \cite{ae1}; the Rabi frequency
of oscillation for a given atomic transition acted upon by a light field 
\cite{af}; nonclassical statistical distributions of light fields \cite{ag},
antibunching effect \cite{ah}; squeezed states \cite{ai,ai1}, and others.

An alternative model that maintains various characteristics of the JCM
and offers advantages in certain situations was proposed by Buck-Sukumar in
1981, abbreviated as BSM \cite{aj}. It is called intensity-dependent JCM,
since it substitutes the JCM interaction  $\lambda (\hat{\sigma}_{+}\hat{a}+%
\hat{\sigma}_{-}\hat{a}^{\dag })$ by another interaction that includes the
number operator $\hat{n},$\ in this way: $ \lambda (\hat{\sigma}_{+}\hat{R}+%
\hat{\sigma}_{-}\hat{R}^{\dag })$ with $\hat{R}=\hat{a}\sqrt{\hat{n}}$ and $%
\hat{R}^{\dag }=\sqrt{\hat{n}}\hat{a}^{\dag }.$ In the previous expressions $\hat{a}$ ($\hat{a}^{\dag }$) stands for annihilation (creation) operator, $\sigma_{-}$ ($\sigma_{+}$) is lowering (raising) operator, ($\hat{n}=\hat{a}^{\dag } \hat{a}$) is the number operator, and $\lambda$ stands for the atom-field coupling. This model also leads to
analytical solution of the Schr\"{o}dinger equation.\ It has been argued
that its physical simulation in laboratory could be implemented via matrices
of waveguides \cite{ak}; optical\ analogies of quantum systems realized in
waveguide arrays have recently impacted the field of integrated optical
structures \cite{gan}. In particular, SUSY photonic lattices can be used to
provide phase matching conditions between large number of modes allowing the
pairing of isospectral crystals \cite{miri,lon,zun,RLM}. In spite of its
apparent theoretical nature the BSM has attracted the attention of various
researchers in the quantum optical community. \cite%
{Shanta,Sivakumar,Ahmed,RLM,Yang,Cardimona,yang1,yang2,Dukelsky,Reeta,Buzek,Valverde,Sayed,Cordeiro}.

In 1992 P. Shanta, S. Chaturvedi, and V. Srinivasana (SCS-model) proposed an
extension of the intensity-dependent JCM \cite{Shanta}. This model
interpolates between the JCM and the BSM . In this approach the authors
assumed the modified Hamiltonian,

\begin{equation}
H_{1}=\omega \hat{N}^{\prime }+\frac{1}{2}\omega _{0}\hat{\sigma}%
_{z}+\lambda (\hat{\sigma}_{+}\sqrt{\hat{N}^{\prime }+1}\hat{a}+\hat{\sigma}%
_{-}\hat{a}^{\dag }\sqrt{\hat{N}^{\prime }+1}),  \label{s1}
\end{equation}%
where $\hat{N}^{\prime }$ is the number operator and the operators $\hat{a}$%
, $\hat{a}^{\dag }$ are quons operators satisfying the the commutation
relation $\hat{a}\hat{a}^{\dag }-q\hat{a}^{\dag }\hat{a}=1$; $q$ is a
c-number restricted to the interval $q\in \lbrack 1,-1].$\ Accordingly,
quons would stand for particles intermediate between bosons ($q=1$) and
fermions ($q=-1$). The authors then use specific connections between the
operator $\hat{N}^{\prime }$ and $\hat{a}$ and $\hat{a}^{\dag }$ and prove
that the SCS model interpolates between the BSM and JCM in the limits $q=1$
and $q=0,$ respectively, with $q$ playing the role of the interpolating
parameter. However, although being a creative approach, here we will not
take it forward because we are restricting ourselves to photonic field, not
to quons. According to Ref. \cite{Dukelsky} there are other nonlinear models
in this context, but they treats the coupled system only approximately \cite%
{vad}

Another type of intensity-dependent JCM, was proposed in 2002 by S.
Sivakumar \cite{Sivakumar}, named here as Sivakumar model (SM).\ This model
also \ interpolates between the JCM and the BSM via the following
Hamiltonian,

\begin{equation}
H=\omega \hat{K}^{\dagger }\hat{K}+\frac{1}{2}\omega _{0}\hat{\sigma}%
_{z}+\lambda (\hat{\sigma}_{+}\hat{K}+\hat{K}^{\dagger }\hat{\sigma}_{-}),
\label{asm1}
\end{equation}%
where $\hat{K}=\sqrt{1+k\hat{a}^{\dagger }\hat{a}}\hat{a}$ and $\hat{K}%
^{\dagger }=\hat{a}^{\dagger }\sqrt{1+k\hat{a}^{\dagger }\hat{a}}$ stand
respectively for annihilation $(\hat{a})$ and creation $%
\left( \hat{a}^{\dagger }\right) $ operators. The change from $\hat{a}$ to $%
\hat{K}$\ aims to get a convenient deformed algebra for various theoretical
applications, as in group theory, field theory, and others. As established in 
\cite{Sivakumar,Ahmed}, for $k=0$ one has the Heisenberg-Weyl algebra
generated by $\{\hat{a},\hat{a}^{\dagger },\hat{I}\}$ and for $k=1$ one
finds the $SU(1,1)$ algebra. For\emph{\ }all values of $k$ the algebra is
closed,

\begin{equation}
\lbrack \hat{K},\hat{K}^{\dagger }]=2\hat{K}_{0},\text{ }[\hat{K}_{0},\hat{K}%
^{\dagger }]=k\hat{K}^{\dagger },\text{ }[\hat{K}_{0},\hat{K}]=-k\hat{K},
\label{asm2}
\end{equation}%
with $\hat{K}_{0}=\ k\hat{a}^{\dagger }\hat{a}+\frac{1}{2}.$ We note some
resemblance between the Hamiltonian in Eq. (\ref{asm1}) and that given by
the BSM for $k=1.$ As pointed out by the authors, the BSM is only reached
when the mean photon number of the field satisfies the condition $k\langle 
\hat{a}^{\dagger }\hat{a}\rangle $ $>>1,$ leading the term $\sqrt{1+k\hat{a}%
^{\dagger }\hat{a}}$ to an approximate form of BSM $\sqrt{\hat{n}},$ $\hat{n}%
=\hat{a}^{\dagger }\hat{a}.$

A somewhat `similar' model, also intensity-dependent, was proposed in 2014
by Rodr\'{\i}gues-Lara \cite{RLM}, named here as Rodr\'{\i}gues-Lara model
(RLM), constituting a generalization of BSM since it substitutes the
operator $\hat{R}=\hat{a}\sqrt{n}$ of the BSM by the operator $\hat{R}=\hat{a%
}\sqrt{\hat{n}+2k}$. The RLM recovers the BSM in the limit $k\rightarrow 0,$
but it includes the counter-rotating terms, due to the form of the
interaction Hamiltonian,

\begin{equation}
H_{int}=\lambda (\sqrt{\hat{n}+2k}\hat{a}+\hat{a}^{\dagger }\sqrt{\hat{n}+2k}%
)\hat{\sigma}_{x},  \label{H-int}
\end{equation}%
where the decomposition $\hat{\sigma}_{x}=\hat{\sigma}_{+}+\hat{\sigma}_{-}$
explains the appearance of counter-rotating terms $\hat{a}^{\dagger }\hat{%
\sigma}_{+}$ and $\hat{a}\hat{\sigma}_{-}.$ As well known, separately they
do not conserve energy. Also, due to the inclusion of the counter rotating
terms, this model puts a restriction on the average number of photons.

In this report we present a generalized Hamiltonian that provides a
continuous and exact interpolation between various Hamiltonian models,
including the JCM, BSM, SM, and RLM. The plan of the paper is as follows. In
Sec. \ref{sec-model}\ we briefly discuss this class of Hamiltonian, showing
its interpolating property. In Sec. \ref{basic} we obtain the solution of
the Schr\"{o}dinger equation in this extended scenario. In Sec. \ref%
{sec-apli} we give some applications, in the Sec. \ref{c1} we calculate
Mandel parameter. The Sec. \ref{cc} contains comments and the conclusion.

\section{Intensity-Dependent Coupling Model\ Hamiltonian}

\label{sec-model}

The Hamiltonian described by the JCM, widely\ referred to as the JCM in the
rotating wave approximation, is given in the form, 
\begin{equation}
\hat{H}=\omega \hat{a}^{\dagger }\hat{a}+\frac{1}{2}\omega _{0}\hat{\sigma}%
_{z}+\lambda _{0}(\hat{\sigma}_{+}\hat{a}+\hat{a}^{\dagger }\hat{\sigma}%
_{-}),  \label{H}
\end{equation}%
where $\omega $ stands for the field frequency, $\omega _{0}$ is the atomic
frequency, and $\lambda $ stands for atom-field coupling. Now, our mentioned
class of interpolating Hamiltonians is obtained substituting $\hat{H}$ by $%
\hat{\mathscr{H}},$ given by

\begin{equation}
\hat{\mathscr{H}}=\omega \hat{a}^{\dagger }\hat{a}+\frac{1}{2}\omega _{0}%
\hat{\sigma}_{z}+\lambda (\hat{\sigma}_{+}\hat{R}+\hat{\sigma}_{-}\hat{R}%
^{\dagger }).  \label{a1a}
\end{equation}
where $\hat{R}=\hat{a}\sqrt{\xi \hat{n}+\delta }$ and $\hat{R}^{\dagger }=%
\sqrt{\xi \hat{n}+\delta }\hat{a}^{\dagger },$ for $\xi \geq 0$ and $0\leq
\delta \leq 1.$

Here it is easily seen that the Hamiltonian in Eq.(\ref{a1a}) interpolates
between the various interaction models of Hamiltonians, as follows:

\begin{itemize}
\item the Jaynes-Cummings model (JCM) \cite{JCM} \ for $\xi =0$ and $%
\delta =1,$

\item the Buck-Sukumar model (BSM) \cite{aj} for $\xi =1$ and $\delta =0,$

\item the Sivakumar model (SM)\ \cite{Sivakumar} for $\xi =k$ and $\delta =1,$

\item  the Rodr\'{\i}gues-Lara model \ (RLM) \cite{RLM} for $\xi =1$ and $\delta
=2k.$
\end{itemize}

Some basic properties involving these atomic and field operators are,

\begin{eqnarray}
\lbrack \hat{\sigma}_{z},\hat{\sigma}_{\pm }] &=&\pm 2\hat{\sigma}_{\pm },%
\text{\ }[\hat{\sigma}_{+},\hat{\sigma}_{-}]=\hat{\sigma}_{z},  \label{cc1}
\\
\lbrack \hat{a},\hat{a}^{\dagger }] &=&1,\text{ }[\hat{a},\hat{n}]=\hat{a},%
\text{ }[\hat{a}^{\dagger },\hat{n}]=-\hat{a}^{\dagger },
\end{eqnarray}%
\begin{equation}
\lbrack \hat{R},\hat{n}]=\hat{R},\text{ }[\hat{R}^{\dagger },\hat{n}]=-\hat{R%
}^{\dagger },\text{ }[\hat{R},\hat{R}^{\dagger }]=2\hat{R}_{0}=\delta +\xi
+2\xi \hat{n},  \label{cc3}
\end{equation}%
with $\hat{R}_{0}=\frac{\delta +\xi }{2}+\xi \hat{n};$ thus we have a closed
algebra in this scenario, 
\begin{equation}
\lbrack \hat{R},\hat{R}^{\dagger }]=2\hat{R}_{0},\text{ }[\hat{R}_{0},\hat{R}%
^{\dagger }]=\xi \hat{R}^{\dagger },\text{ }[\hat{R}_{0},\hat{R}]=-\xi \hat{R%
}.
\end{equation}

The Eq.(\ref{a1a}) can be rewritten in the form,

\begin{equation*}
\hat{\mathscr{H}}=\hat{\mathscr{H}}_{A}+\hat{\mathscr{H}}_{I}\text{ },
\end{equation*}%
where,

\begin{eqnarray}
\hat{\mathscr{H}}_{A} &=&\omega (\hat{a}^{\dagger }\hat{a}+\frac{1}{2}\hat{\sigma}%
_{z}), \\
\hat{\mathscr{H}}_{I} &=&\frac{1}{2}\Delta \omega \hat{\sigma}_{z}+\lambda (%
\hat{\sigma}_{+}\hat{R}+\hat{\sigma}_{-}\hat{R}^{\dagger }),  \label{cv2}
\end{eqnarray}%
with $\Delta \omega =$\ $\omega _{0}-\omega .$\ Next we can use the Eqs.(\ref%
{cc1}) and (\ref{cc3}) to show that $\hat{\mathscr{H}}_{A}$ and $\hat{%
\mathscr{H}}_{I}$ \ are constant of motion, namely, 
\begin{equation}
\lbrack \hat{\mathscr{H}},\hat{\mathscr{H}}_{A}]=[\mathscr{H},\hat{%
\mathscr{H}}_{I}]=[\hat{\mathscr{H}}_{A},\hat{\mathscr{H}}_{I}]=0.
\end{equation}%
All essential dynamic properties contained in a state of the atom-field
system described by any of the previous interpolating Hamiltonians, can also
be described by the interpolating Hamiltonian proposed here, $\hat{%
\mathscr{H}}_{I}$, considering that $\hat{\mathscr{H}}_{A}$ contributes only
for general phase factors, usually not relevant.

\section{Field fluctuations}

\label{basic}

Let us consider a simple example assuming the system in resonance, $\Delta
\omega =0$%
\begin{equation}
\hat{\mathscr{H}_{I}}=\lambda (\hat{\sigma}_{+}\hat{R}+\hat{\sigma}_{-}\hat{R%
}^{\dagger }).  \label{ac}
\end{equation}%
Now, to analyze the time evolution of the coupled atom-field system we solve
the time dependent Schr\"{o}dinger equation using the Hamiltonian in Eq.(\ref%
{ac}),

\begin{equation}
i\frac{d|\Psi (t)\rangle }{dt}=\hat{\mathscr{H}_{I}}|\Psi (t)\rangle .
\label{a3a}
\end{equation}%
We can write the formal solution of Eq. (\ref{a3a}) as,

\begin{equation}
|\Psi (t)\rangle =\hat{U}(t)|\Psi (0)\rangle ,
\end{equation}%
where $\hat{U}(t)=\exp (-i\hat{\mathscr{H}_{I}}t)$, is the (unitary)
evolution operator. \ Next, using the expression 
\begin{equation}
e^{-\beta \hat{u}}\equiv \sum\limits_{n=0}^{\infty }\frac{(-\beta )^{n}}{n!}%
\hat{u}^{n},
\end{equation}%
and decomposing the above sum in their even and odd terms, plus the use of
the two following relations

\begin{widetext}
\begin{equation}
\hat{U}(t)=\sum\limits_{n=0}^{\infty }\frac{(-1)^{n}(\lambda t)^{2n}}{(2n)!}(%
\hat{\sigma}_{+}\hat{R}+\hat{\sigma}_{-}\hat{R}^{\dagger
})^{2n}-i\sum\limits_{n=0}^{\infty }\frac{(-1)^{n}(\lambda t)^{2n+1}}{(2n+1)!%
}(\hat{\sigma}_{+}\hat{R}+\hat{\sigma}_{-}\hat{R}^{\dagger })^{2n+1},
\label{cc5}
\end{equation}

\end{widetext} 
and,
\begin{equation}
(\hat{\sigma}_{+}\hat{R}+\hat{\sigma}_{-}\hat{R}^{\dagger })^{2n}=\left( 
\begin{array}{cc}
\sqrt{\hat{R}\hat{R}^{\dagger }}^{2n} & 0 \\ 
0 & \sqrt{\hat{R}^{\dagger }\hat{R}}^{2n}%
\end{array}%
\right) ,
\end{equation}%
with $(\hat{R}\hat{R}^{\dagger })^{N}\hat{R}=\hat{R}(\hat{R}^{\dagger }\hat{R%
})^{N}$, we get the evolution operator in a convenient form for systems
involving a two-level atom,%
\begin{eqnarray}
\hat{U}(t) &=&\cos (\lambda t\sqrt{\hat{A}+2\hat{R}_{0}})|e\rangle \langle e|%
\text{ }+\cos (\lambda t\sqrt{\hat{A}})|g\rangle \langle g|  \notag \\
&&-i\hat{R}\frac{\sin (\lambda t\sqrt{\hat{A}})}{\sqrt{\hat{A}}}|e\rangle
\langle g|\text{ }-i\frac{\sin (\lambda t\sqrt{\hat{A}})}{\sqrt{\hat{A}}}%
\hat{R}^{\dagger }|g\rangle \langle e|,  \notag \\
\label{alfa}
\end{eqnarray}%
with $\hat{A}=\hat{R}^{\dagger }\hat{R},$ and $\hat{R}$, $\hat{R}^{\dagger }$
given above.

We will assume the entire system initially decoupled, $|\Psi (0)\rangle
=|\psi \rangle |g\rangle ,$ the atom in its ground state $|g\rangle $ and
the field in arbitrary state $|\psi \rangle .$ So, the wavefunction
describing the atom-field system for arbitrary times is obtained from
equation $|\Psi (t)\rangle =U(t)|\Psi (0)\rangle $ with $U(t)$ given in Eq. (%
\ref{alfa}). After an algebraic procedure we find,

\begin{equation}
|\Psi (t)\rangle =\cos (\lambda t\sqrt{\hat{A}})|\psi \rangle |g\rangle -i%
\hat{R}\frac{\sin (\lambda t\sqrt{\hat{A}})}{\sqrt{\hat{A}}}|\psi \rangle
|e\rangle .  \label{a2a}
\end{equation}

\section{Paradoxical evolution of average number of photons\label{sec-apli}}

The paradox concerned with the time evolution of the average number of
photons, discussed by Luis \cite{luis}, used the JCM. Here we treat this
paradox for the various interpolating Hamiltonians mentioned above. This is
obtained directly from our Hamiltonian by an appropriate choice of the pair $%
\xi $ and $\delta $.

The mean number of photons of the field is calculated as,%
\begin{equation}
\langle \hat{n}\rangle =Tr(\hat{n}\hat{\rho})=\sum_{n=0}^{\infty }\langle n|%
\hat{n}\hat{\rho}|n\rangle ,
\end{equation}%
where $\hat{\rho}=|\psi \rangle \langle \psi |$\emph{\ }is the density
operator.

In this section we study the dynamic behavior of the average number of
photons, $\langle \hat{n}(t)\rangle =\langle \Psi (t)|\hat{n}|\Psi
(t)\rangle ,$ $\langle \hat{n}(t)\rangle _{g}=\langle \psi _{g}|\hat{n}|\psi
_{g}\rangle $ and $\langle \hat{n}(t)\rangle _{e}=\langle \psi _{e}|\hat{n}%
|\psi _{e}\rangle ,$ where

\begin{eqnarray}
|\psi _{g}\rangle &=&\cos (\lambda t\sqrt{\hat{A}})|\psi \rangle |g\rangle ,
\\
|\psi _{e}\rangle &=&-i\hat{R}\frac{\sin (\lambda t\sqrt{\hat{A}})}{\sqrt{%
\hat{A}}}|\psi \rangle |e\rangle .
\end{eqnarray}%
\ For small times the following average values are obtained,

\begin{equation}
\langle \hat{n}(t)\rangle _{g}=\langle \hat{n}\rangle -\lambda ^{2}t^{2}[
\xi ( \langle \hat{n}^{3}\rangle -\langle \hat{n}^{2}\rangle \langle \hat{n}%
\rangle ) +\delta \Delta \hat{n}^{2}] ,  \label{ng}
\end{equation}

\begin{equation}
\langle \hat{n}(t)\rangle _{e}=\frac{\langle \hat{n}\hat{A}\rangle }{\langle 
\hat{A}\rangle }-1-\frac{\lambda ^{2}t^{2}}{3\langle \hat{A}\rangle ^{2}}%
[\langle \hat{n}\hat{A}^{2}\rangle \langle \hat{A}\rangle -\langle \hat{n}%
\hat{A}\rangle \langle \hat{A}^{2}\rangle ],  \label{ne}
\end{equation}

\begin{equation}
\langle \hat{n}(t)\rangle =\langle \hat{n}\rangle (1-\lambda ^{2}t^{2}\delta
)-\lambda ^{2}t^{2}\xi \langle \hat{n}^{2}\rangle ,  \label{n}
\end{equation}%
where $\langle \hat{A}\rangle =\xi \langle \hat{n}^{2}\rangle +\delta
\langle \hat{n}\rangle ,$ $\langle \hat{n}\hat{A}\rangle =\xi \langle \hat{n}%
^{3}\rangle +\delta \langle \hat{n}^{2}\rangle ,$ $\langle \hat{n}\hat{A}%
^{2}\rangle =\xi ^{2}\langle \hat{n}^{5}\rangle +2\xi \delta \langle \hat{n}%
^{4}\rangle +\delta ^{2}\langle \hat{n}^{3}\rangle ,$ $\langle \hat{A}%
^{2}\rangle =\xi ^{2}\langle \hat{n}^{4}\rangle +2\xi \delta \langle \hat{n}%
^{3}\rangle +\delta ^{2}\langle \hat{n}^{2}\rangle .$

Regardless of the types of interpolations, i.e., JCM $\longleftrightarrow $
BSM$\longleftrightarrow $ SM $\longleftrightarrow $ RLM, and
eventually others obtained by varying the pair $\xi $\ and $\delta $\emph{, }%
the essential features of the paradox discussed in Ref. \cite{luis} remains for
all these interpolation models. Now, for small times, the following relation
is valid, irrespective of the interpolating model. 
\begin{equation}
\langle \hat{n}(t)\rangle _{e}>\langle \hat{n}(t)\rangle >\langle \hat{n}%
(t)\rangle _{g}.
\end{equation}

\begin{figure}[h!tb]
\centering  
\includegraphics[width=8cm, height=7cm]{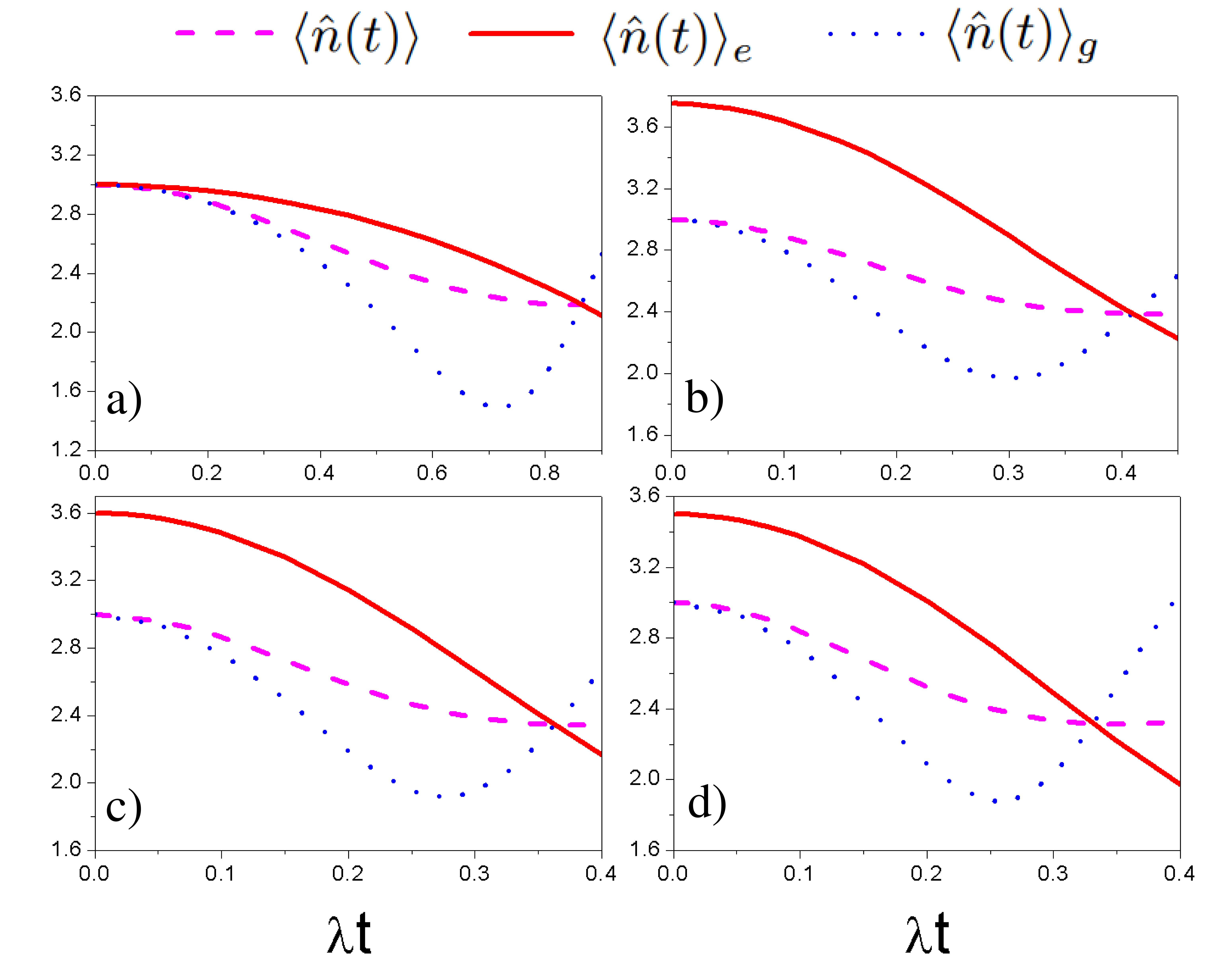}
\caption{Evolution of $\langle \hat{n}(t)\rangle _{e}$ (solid curve) $%
\langle \hat{n}(t)\rangle $ (dashed curve) and $\langle \hat{n}(t)\rangle
_{g}$ (dotted curve), for an initial coherent state with $\langle \hat{n}%
\rangle =3;$ a) JCM, b) BSM, c) SM and d) RLM.}
\label{figura_1}
\end{figure}

In the plots of Fig. (\ref{figura_1}) we have assumed the initial field in a
coherent state, assuming the average number of photons $\langle \hat{n}%
\rangle =3.$ Here we have used mathematical expressions more general
than those in Eqs. (\ref{ng}, \ref{ne} and \ref{n}), hence the following
plots are not restricted to small times\emph{. }We observe in Fig. (\ref%
{figura_1} (a)) the occurrence of the mentioned paradox, which starts
immediately and remains up to $\lambda t\approx 0.7$ for the JCM; \ in Fig. (%
\ref{figura_1} (b)) $\ \lambda t\approx 0.3$ for the BSM; in Fig. (\ref%
{figura_1} (c)) $\lambda t\approx 0.27$ for the SM; and in Fig. (\ref%
{figura_1} (d)) \ $\lambda t\approx 0.25$ for the RLM.

Hence, these results show that the paradox raised by\emph{\ }A. Luis \cite%
{luis} using the JCM happens no matter what kind of Hamiltonian model used
within the class considered here.

\section{Estat\'{\i}stica Sub-Poissoniana\label{c1}}

A quantized photon field with sub-Poissonian statistics is characterized
when the variance is smaller than the average number of photons, namely: $%
\langle \Delta \hat{n}^{2}\rangle =(\langle \hat{n}^{2}\rangle -$ $\langle 
\hat{n}\rangle ^{2})<\langle \hat{n}\rangle ;$ the opposite chacterizes a
super-Poissonian photon field and if $\langle \Delta \hat{n}^{2}\rangle
=\langle \hat{n}\rangle $ the photon field exhibits Poissonian statistics,
characterizing all coherent states. The Mandel's parameter tells us what
kind of statistics the field displays \cite{mandel}; it is given by the
relation,%
\begin{equation}
Q=\frac{\langle \Delta \hat{n}^{2}\rangle }{\langle \hat{n}\rangle }-1.
\end{equation}

So, when $Q>0$ the field is super-Poissonian; when $Q<0$ it is
sub-Poissonian; and Poissonian for $Q=0$.

\begin{figure}[h!tb]
\centering  
\includegraphics[width=8cm, height=7cm]{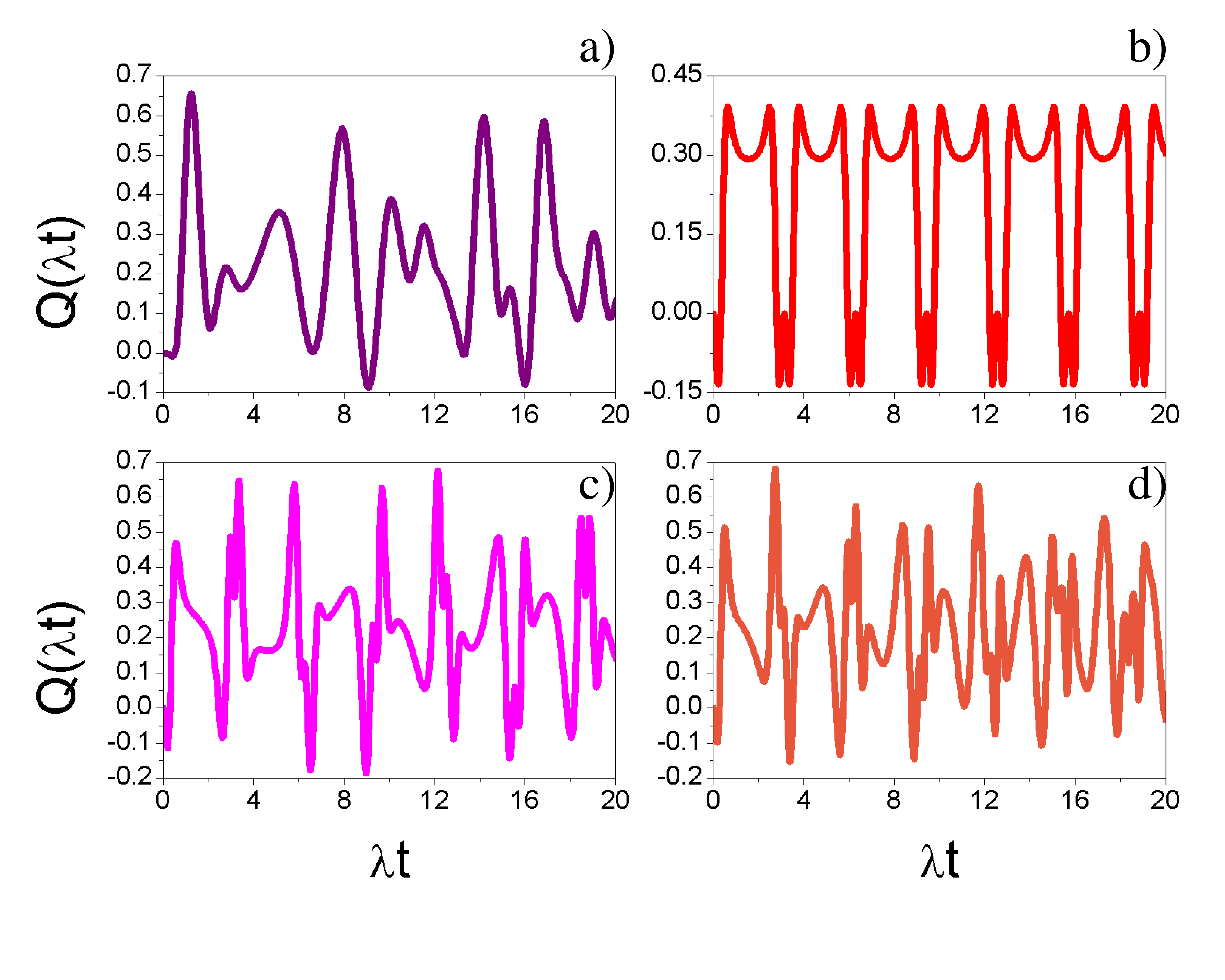}
\caption{Evolution of $Q(\protect\lambda t)$, for an initial coherent state
with $\langle \hat{n}\rangle =3;$ a) JCM, b) BSM, c) SM and d) RLM.}
\label{figura_2}
\end{figure}

In Fig. (\ref{figura_2}), we represent our Hamiltonian in Eq. (\ref{cv2})
interpolating between the four Hamiltonians: JCM, BSM, SM and RLM. 

\begin{figure}[h!tb]
\centering  
\includegraphics[width=.50\textwidth]{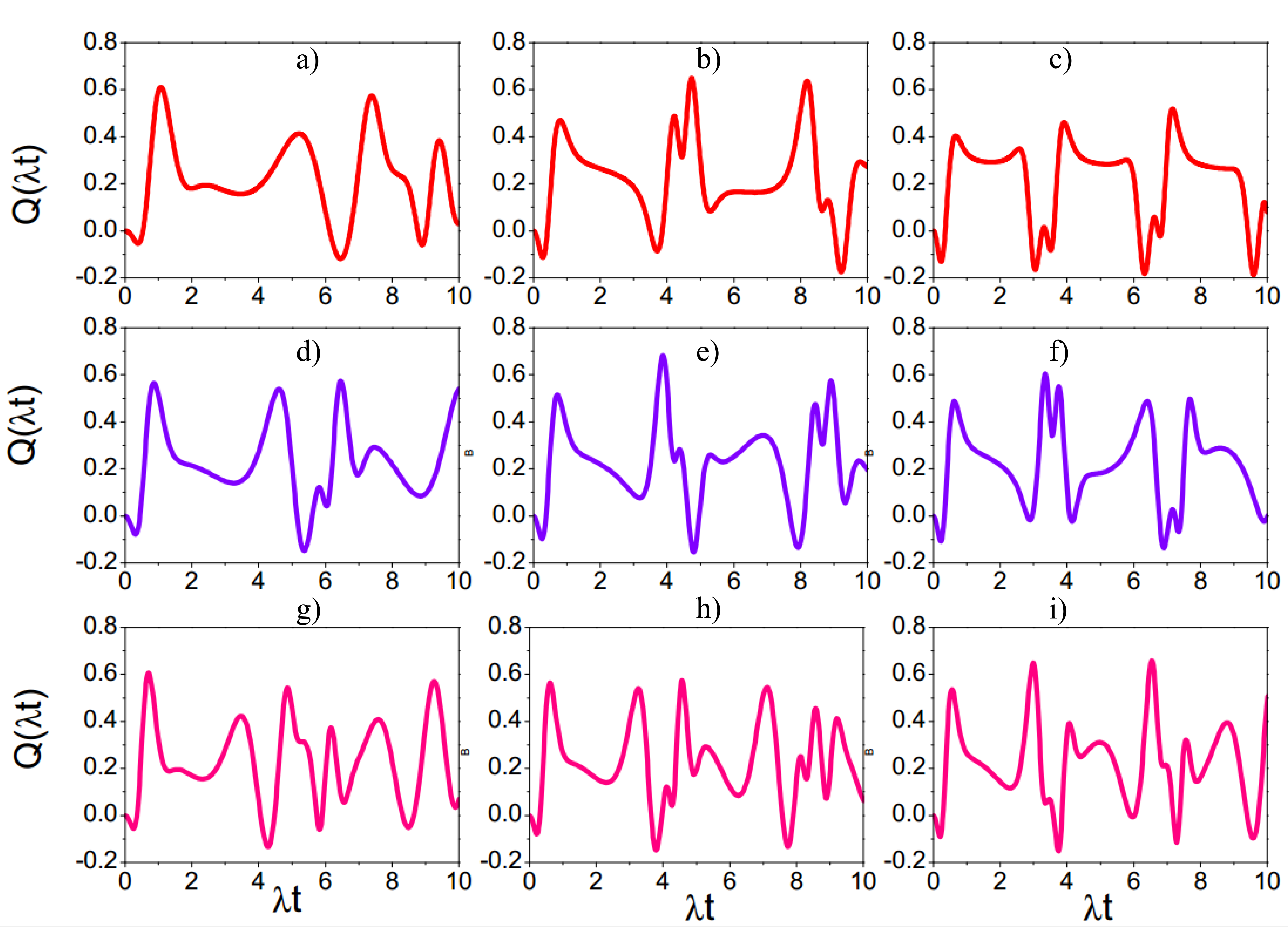}
\caption{Evolution of $Q(\protect\lambda t)$ for an initial coherent state
with $\langle \hat{n}\rangle =3$; interpolating from JCM to BSM a) for $\protect\xi %
=0.10$ and $\protect\delta =0.90;$ b) for $\protect\xi =0.50$ and $\protect%
\delta =0.50;$ c) for $\protect\xi =0.90$ and $\protect\delta =0.10$;
interpolating from JCM to SM d) for $\protect\xi =0.25$ and $\protect\delta %
=1; $ e) for $\protect\xi =0.50$ and $\protect\delta =1$; f) for $\protect%
\xi =0.75$ and $\protect\delta =1$; interpolating from JCM to RLM g) for $%
\protect\xi =0.25$ and $\protect\delta =2$; h) for $\protect\xi =0.50$ and $%
\protect\delta =2;$ i) for $\protect\xi =0.75$ and $\protect\delta =2.$}
\label{figura_3}
\end{figure}

Fig.(\ref{figura_3}) exhibits various plots of the Mandel parameters in
these different models of Hamiltonian. The various plots show that, by
conveniently adjusting the pair of parameters $\xi $ and $\delta $\ in the
present model Hamiltonian we can interpolate continuously from the JCM to
the BSM, the SM, and the RLM. In these interpolations we have observed in
which way the Mandel parameter modifies during the time evolutions, as shown
in Fig.(\ref{figura_3}), plots (a), (b), and (c); also, this interpolation
occurs in a softly way, from the JCM to BSM. The same happens for the
interpolation from the JCM to the SM, shown in Fig.(\ref{figura_3}), plots
(d), (e), and (f); and also from the JCM to the RLM, Fig.(\ref{figura_3}),
plots (g), (h), and (i).


\begin{figure}[h!tb]
\centering  
\includegraphics[width=.50\textwidth]{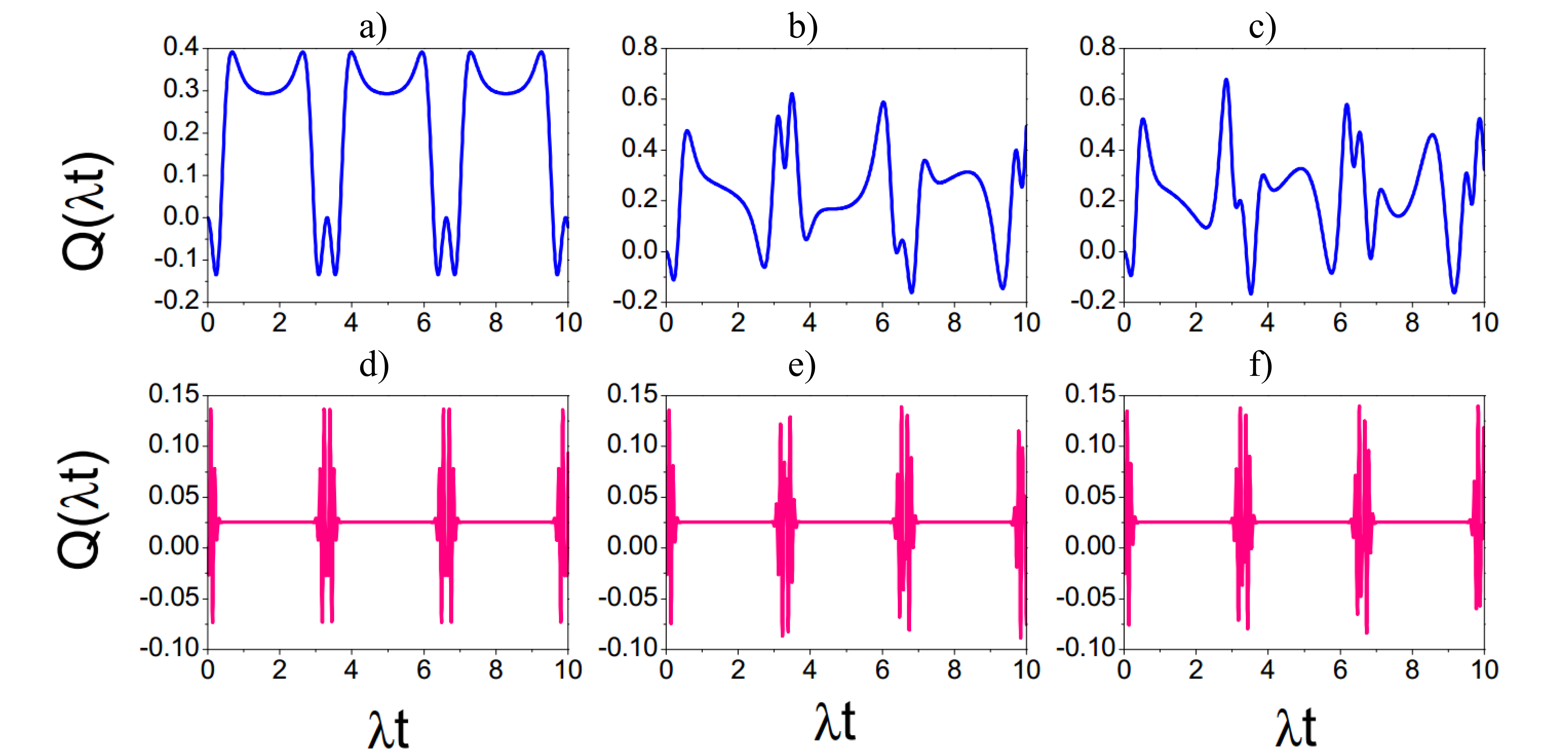}
\caption{Evolution of $Q(\protect\lambda t)$, for an initial coherent state;
a) with $\langle \hat{n}\rangle =3;$ para $\protect\xi =0.90$ and $\protect%
\delta =0;$ b) with $\langle \hat{n}\rangle =3;$ para $\protect\xi =0.90$
and $\protect\delta =1;$ c) with $\langle \hat{n}\rangle =3;$ para $\protect%
\xi =0.90$ and $\protect\delta =2;$ d) with $\langle \hat{n}\rangle =30;$
para $\protect\xi =0.90$ and $\protect\delta =0;$ e) with $\langle \hat{n}%
\rangle =30;$ para $\protect\xi =0.90$ and $\protect\delta =1;$ f) with $%
\langle \hat{n}\rangle =30;$ para $\protect\xi =0.90$ and $\protect\delta %
=2; $ }
\label{Figura_4}
\end{figure}
We can note in Fig.(\ref{Figura_4}) that, when we compare the case where the
system state has a small average excitation $\langle \hat{n}\rangle $ with
those having larger values of $\langle \hat{n}\rangle $, the Mandel
parameters for different Hamiltonians differ sensitively from each other for
small values of\ $\langle \hat{n}\rangle $, the region where the quantum
nature of the system state is more evident. Contrarily, for larger values $%
\langle \hat{n}\rangle $ the corresponding plots are very similar. In theses
examples we are analizing the Mandel parameter close to BSM, Fig.(\ref%
{Figura_4} a) and d)), with other close to SM, Fig.(\ref{Figura_4} b), and
e)), and another close to RLM, Fig.(\ref{Figura_4} c) and f)). This shows a
 great sensitivity of the system to the
parameters $\xi $ and $\delta $ in the quantum regime of small numbers, as
usually expected. In addition, for small values of $\langle \hat{n}\rangle $
 the field state exhibts a greater sub-Poissonian effect.
 \newline
\section{Conclusion}
\label{cc}

We have proposed a (two parameters)\ interpolating Hamiltonian. It allows
one to extend from (a) the JCM, \ (b) the BSM, (c) the SM, \ and (d) the
RLM. This new Hamiltonian employs the basic operators $\hat{R}=\hat{a}\sqrt{%
\xi \hat{n}+\delta }$, $\hat{R}^{\dagger }=\sqrt{\xi \hat{n}+\delta }\hat{a}%
^{\dagger }$\ , and $\hat{R}_{0}=\frac{\delta +\xi }{2}+\xi \hat{n}$ which
form a closed algebra. As mentioned before, it contains all essential
dynamic properties contained in a state of the atom-field system described
by the previous interpolating Hamiltonians. To give an example we have
verified that, essentially, the results found in the paradox discussed by A.
Luis \cite{luis} in the JCM remains in the scenario of this extended Hamiltonian
(see Fig. (\ref{figura_1})), no matter the chosen extension, say: from (a)
to (b), from (a) to (b), from (a) to (c), and from (a) to (d). We have also
calculated the Mandel parameter to obtain the evolution of the statistical
properties of the system state and their time evolution when we pass from
our interpolating model to another after appropriate choices of the pair $%
\xi ,\delta .$\ In these time evolutions we have highlighted the influence
of the average excitation $\langle \hat{n}\rangle ,$\ when large or small,
upon the statistical properties of the system. From what we have
learned in quantum optics, concerning the degradation caused by decoherence
effects affecting quantum states \cite{haroche}, for practical purposes this
result would lead us to give priority to states with smaller excitations,
the quantum region of small numbers, where some types of interpolating
Hamiltonians have problems \cite{exc}.

\section{Acknowledgements}

We thank the Brazilian funding agencies CNPq and FAPEG for the partial
supports.

\end{document}